\documentclass[journal=langmuir,manuscript=article]{achemso}

\usepackage[T1]{fontenc} 
\usepackage{multirow}

\author{K Nilavarasi}
\affiliation{Department of Physics, School of Basic and Applied Sciences, Central University of Tamil Nadu, Thiruvarur-610005, Tamil Nadu, India}
\email{nilavarasi@students.cutn.ac.in}
\author{S G Ramkumar}
\affiliation{Department of Chemistry, School of Basic and Applied Sciences, Central University of Tamil Nadu, Thiruvarur-610005, Tamil Nadu, India}
\email{ramkumar@cutn.ac.in}
\author{V Madhurima}
\affiliation{Department of Physics, School of Basic and Applied Sciences, Central University of Tamil Nadu, Thiruvarur-610005, Tamil Nadu, India}
\email{madhurima@cutn.ac.in}

\title[An \textsf{achemso} demo]
  {1D roughness driven depinning of self-assembly of liquid droplets}

\abbreviations{IR,NMR,UV}

\begin{document}



\begin{abstract}
The influence of surface constraints on the self-assembly of liquid droplets is investigated.  A semi-quantitative explanation for large scale pattern formation consisting of small scale closely arranged droplets inside the large scale distorted ring of droplets is presented in this paper. The scale at which the influence of constraints become dominant is also determined in this study.  It is seen that the  underlying roughness has a larger impact than the nature of polymer on pore size. Comparative studies of pore patterns formed on smooth and constrained substrates are reported. The simulated energy minimized shape of the droplets on smooth and constrained substrates are obtained using \textit{Surface Evolver}.

\end{abstract}

\section{Introduction}
The patterns formed from self-assembly due to the condensation of liquid droplets has attracted great attention due to formation of ordered condensates \cite{zlixma, qhong, tran, lswan, hyuanyu, jmansouri, cduzhang, asdleon, maudsave, ypngjian}. If the surface onto which the droplets condensate is free flowing, orderly porous structure appears resulting in a honeycomb structure \cite{mohan}.  However, there are reports on large scale mesoscopic patterns containing both small scale honey comb patterns inside a net-like large scale patterns \cite{edwardborm, edward}. These large scale mesoscopic patterns are formed due to the rapid cooling of the solvent/vapor interface resulting from the evaporation of the solvent \cite{edward}. Conventionally, the patterns are attributed to the surface tension driven Marangoni instability \cite{ mohan, maruyama}. Whereas, Bormashenko et. al., related these large scale mesoscopic patterns to the combined effect of mass transport instability and solvent vapor bubble migration driven by plateau borders \cite{borm2, edward, doronborm}.  The former, mass transport instability, the instability introduced by Pierre de Gennes,  refers to the instability of polymer-rich layer adjacent to the solution/vapor interface \cite{gennes, degennes, borm2} and the latter occurs due to the capillary pressure gradient formed within the plateau borders \cite{doronborm, borm2}.  The mechanism of formation of these large scale patterns still remains unclear. \\

\begin{figure}[h]
\centerline{\includegraphics[width=5in]{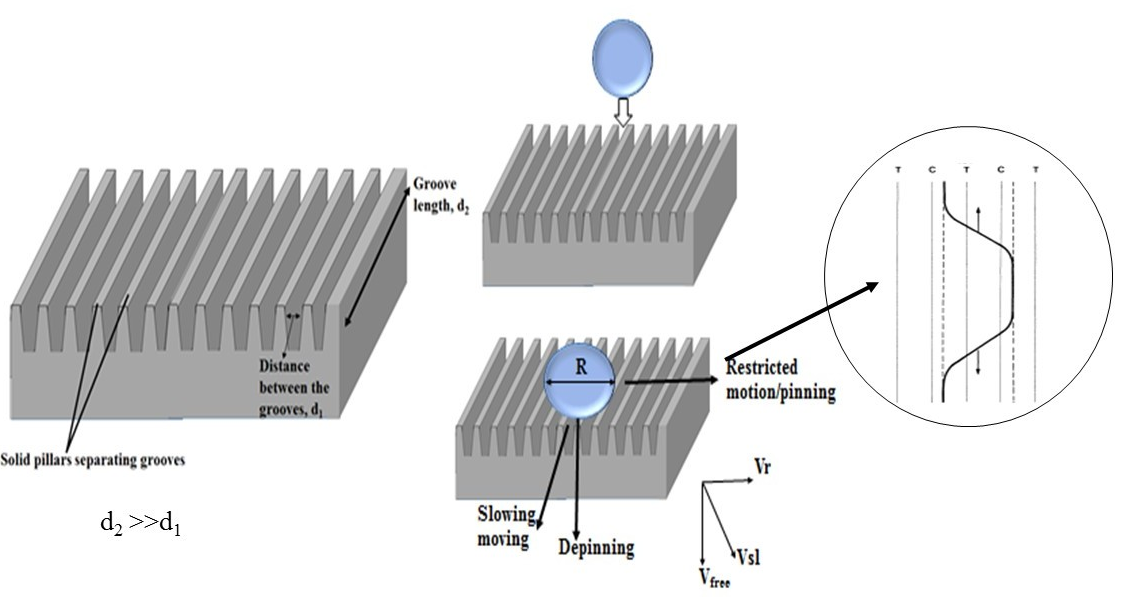}}
\caption{Schematic representation showing the pinning/depinning of triple phase contact line on a parallel grooved surface.}
\label{grooves}
\end{figure}

In the present work, the formation of a pattern very similar to the large scale mesoscopic patterns observed in the literature \cite{edward} is reported.  These patterns are observed on the films patterned on constrained substrates and the patterns are obtained when the drop-casted polymer is allowed to dry in a saturated humid atmosphere. Although various explanations regarding the formation of large scale patterns are available, the mechanism is still not clearly understood \cite{edward}.  In this paper, the pattern formation is related to the pinning and depinning of triple phase contact line of the condensed droplets.  In the seminal paper of de Gennes, it was stated that, when there are parallel grooves of the scale of the order of capillary length, the triple phase contact line can have two different types of movement.  One is the movement of contact line parallel to the grooves and the other is the movement of contact line making an angle $\psi$ to the grooves \cite{gennestpcl}. The schematic representation of the movement of contact line over parallel grooved surface is given in Figure~\ref{grooves}. The  Movement of triple phase contact line is used in the present study to explain the pattern formation and conclude that the interesting patterns result from the movement of triple phase contact line.  It has to be emphasized that the comprehensive understanding of the formation of interesting patterns occurring with the constrained surface has still not been achieved.  \\

It is noted from the literature that if the scale of the underlying constraints is of the order of $50$ $\mu m$, then the constraints have no influence on the self-assembly process \cite{jspark, pierreescale}. To confirm that the underlying constraints is the only influencing parameter, the experiment is performed onto a smooth surface and as expected, a stable, energy minimized honey-comb structure is observed on smooth surfaces. To further confirm that the formation of large scale patterns are independent of the polymer, in this work, the patterning is performed with two different polymers and of varying molecular weight. The energy minimized shape of the droplets is simulated using \textit{Surface Evolver}, a program for the modeling of liquid surfaces shaped by various forces and constraints through an energy minimization \cite{brake, brake2}. These shapes are used to compare the radius of the pores obtained from the experimental study.\\

The present study reports the influence of underlying constraints on the self-assembly of liquid droplets and hence the large scale patterns are related to the pinning and depinning of triple phase contact line in addition to the conventional instabilities.  A semi-quantitative explanation for the formation of patterns on the constrained substrates is also discussed.   \\

\section{Experimental section}
Three polymers, Polydimethyl siloxane (PDMS) of molecular weight $236.53$ $g/mol$, Poly(Bisphenol A carbonate)  of molecular weight $28,200$ $g/mol$ and Poly(Bisphenol A carbonate) of molecular weight $45,000$ $g/mol$ were purchased from SIGMA Aldrich and individually dissolved in chloroform (purity $99.9 \%$) purchased from EMPLURA which is the solvent in the present study.  The concentration of the polymer solution was $5 \%$ weight by volume.  \\

Two different type of substrates with smooth and constrained surface were used and were drop casted with $300$ $\mu l$ of polymer solution and are kept under the saturated atmosphere of water vapor at $26$ $^{o}C$. The details regarding the substrates were clearly given in our recent papers \cite{nilavarasi, materials}. 3 $ml$ of distilled water was added in a petri-dish and kept inside a sealed glass chamber to form a saturated atmosphere of water.\\

The substrates were positioned at least $1$ $cm$ above the water level.  After 15 minutes, (i.e., after the complete evaporation of the solvent), the substrate surface is left with a thin layer of polymer.  The experimental setup and its schematic diagram are as reported previously in Ref.\cite{nilavarasi}. The patterns were studied with confocal laser scanning microscope and the wetting studies were performed with Rame Hart Contact angle goniometer.   \\

\section{Simulation details}
The equilibrium shape of the liquid droplet can be simulated through energy minimization. This is performed using \textit{Surface
Evolver}, an interactive program which uses Finite element method and gradient descent method to evolve the surface toward minimal energy \cite{brake}. The equilibration of the system is achieved by minimizing various energies namely, surface tension, gravitational energy, etc., involved in the defined system. The initial geometrical parameters, energies and constraints involved in the
system are given as inputs and the program minimizes the total energy of the system by modifying the surface geometry according to the defined parameters and constraints. This program is used to simulate the stable equilibrium shape of various liquid
droplets over specific smooth and constrained surfaces and the results are compared with those from experiments.

In this work, the \textit{Surface Evolver} is used to simulate a sessile drop of water onto a smooth and constrained surface. The bottom face of the droplet is constrained to move and this boundary condition is considered to be responsible for obtaining the shape of the droplets. For the sake of convenience, the constraints are specified to the edges which defines the three phase contact line.
The successive refinements and steps concerning energy minimization computes the equilibrium shape of the system. Since the gravitational effect is negligible for the micro-sized condensed droplets, the gravitational energy was not taken into account in the present simulations. For a given values of surface tension of the droplet and the inter-facial energy, the initial geometry of the droplet is evolved to attain the final equilibrium shape.  The free energy of the system $G$ is expressed as,
\begin{equation}
G/\gamma_{lv} = A_{lv} - \int\int \cos \theta dA
\label{eq:}
\end{equation}
and the contact angle is defined by Young's equation \cite{young},
\begin{equation}
\cos \theta = \frac{(\gamma_{SV} - \gamma_{SL})}{\gamma_{LV}},
\label{eq:}
\end{equation}
where $A_{lv}$ is the contact area of the liquid, $\gamma_{lv}$ is the liquid vapor inter-facial tension, $\gamma_{SV}$ is the solid-vapor inter-facial tension, $\gamma_{SL}$ is the solid-liquid inter-facial tension and $\theta$ is the Young's contact angle.
The drop volume specified in this work is $9 \mu l$.  \\

\section{Results and Discussion}
\begin{figure}[h]
\centerline{\includegraphics[width=4in]{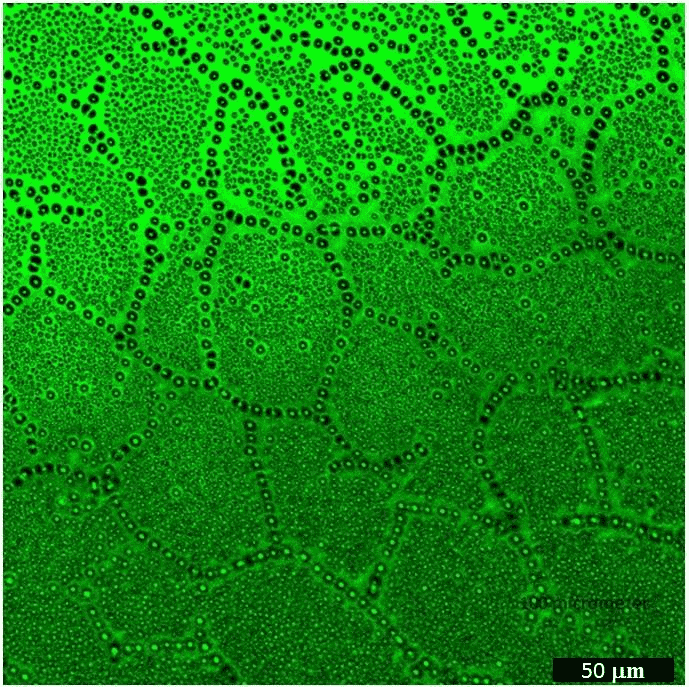}}
\caption{CLSM image of pattern formation on PDMS coated constrained surface.}
\label{waterconstraint}
\end{figure}

Self-assembled droplet pattern formed due to reorganization of PDMS film over the constrained substrate is shown in Figure~\ref{waterconstraint}. The pattern comprises of small scale closely arranged pores within a large scale distorted hexagonal network. The pattern formation can be explained through the underlying principle of moving contact lines, pinning of droplets and the discontinuity of energy minimum on a constrained  surface. Provided the droplet size is larger than the groove width, for a parallel grooved surface with grooves infinitely long and close to each other, there can be movement of contact line which lies at an angle to the grooves \cite{gennestpcl}. This results in the formation of droplets having a local structure due to the continuous displacement of contact line without any pinning \cite{marmur, erbil}.\\

The ring formation is attributed to the moving three-phase contact line and change in its direction due to pinning of droplets at certain points and successive evaporation of droplets. On the constrained surface, after evaporation of certain droplets, there are spaces for occupying new droplets. This favors re-nucleation at the spaces inside the ring. Thus the large scale pore pattern on constrained surface is attributed to the geometry of the underlying surface in addition to the Marangoni effect and plateau border effect \cite{edwardborm}. \\
\begin{figure}[h]
\centerline{\includegraphics[width=5in]{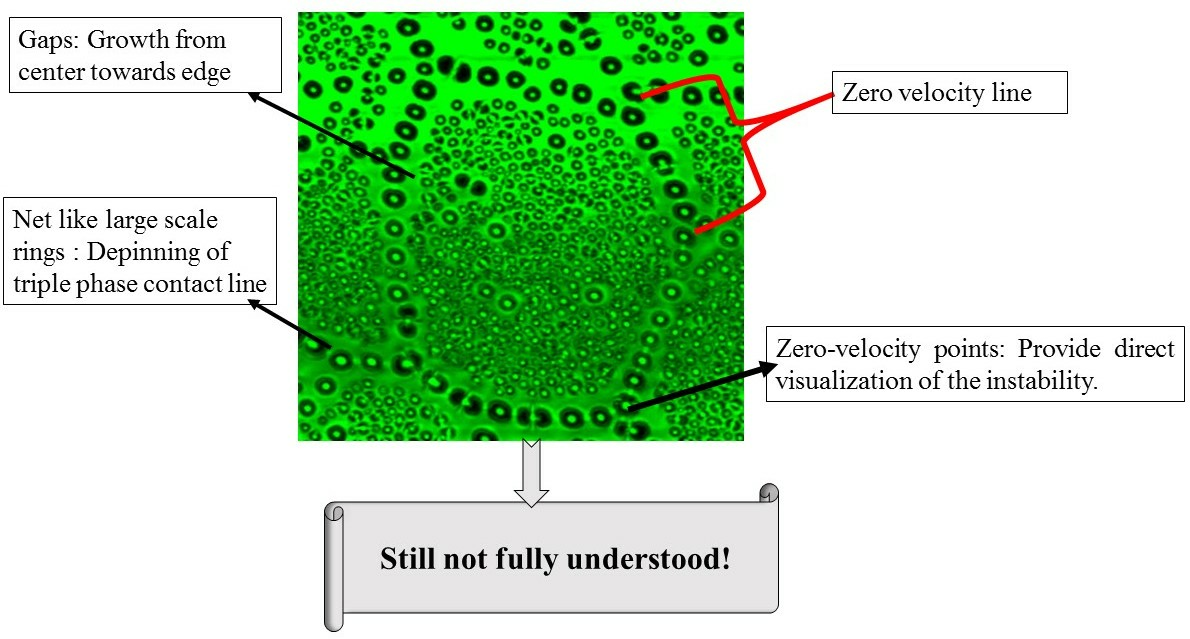}}
\caption{Zoomed portion of a single distorted ring taken from Figure~\ref{waterconstraint}.}
\label{zoomconst}
\end{figure}
As indicated in the theory by de Gennes \cite{gennestpcl}, the argument of contact line moving is valid for the order of length scale of $~1$ $\mu m$ (i.e., capillary length of the liquid) \cite{gennestpcl}. So with one dimensional grooves separated by $0.7$ $\mu m$,  a dual pattern of imprints of small clusters inside a distorted hexagonal ring of droplets is observed. The same pattern as seen above is also observed in non-aqueous environment i.e. with ethanol instead of water and is discussed in detail in our recent papers \cite{nilavarasi, materials}. Whereas with 1D grooves separated by $50$ $\mu m$, dual honey comb structures are formed: one over the groove and other in between the groove \cite{jspark, pierreescale}. These arguments of contact line movement holds along with the conventional argument of instabilities for the pattern formation.  \\

\subsection{Formation of large pore at the corners}
The large pores at the corners of each distorted ring (Figure~\ref{zoomconst}) are related to the plateau border and the zero velocity point as discussed by Bormashenko et.al.,\cite{edward}  Usually, Plateau Border effect is used to describe the structure (shape and configuration) of soap films. Many natural patterns obey this effect.  Plateau border is defined as an edge formed by the intersection of three lamella - a thin film separating the air bubbles within the foam, at an angle of $120^{o}$.  Similar to the migration of soap films, the polymer solution/vapor inter-facial layer migrate to form a plateau border of polymer in an evaporating solution \cite{doronborm, edward}.  This effect is similar to the plume effect, as observed by de Gennnes \cite{gennes, degennes}.  The droplets upon evaporation of solvent moves towards the corner, the zero velocity point, of the distorted rings through the plateau border formed in an evaporating solution. Zero velocity point is defined as a point of any continuous tangent vector film on the sphere at which the velocity is zero.  This is also called as the Hairy ball theorem \cite{maruyama, eisenberg}.   

\subsection{Influence of polymer}
\begin{figure}[h]
\centerline{\includegraphics[width=6in]{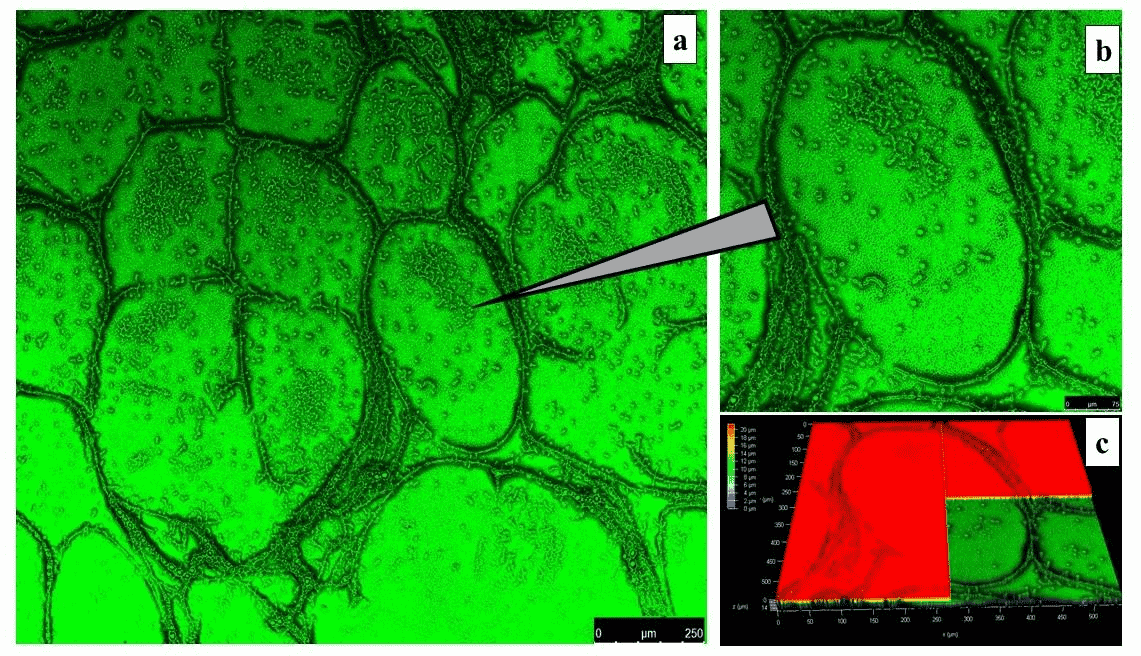}}
\caption{CLSM image of pattern formation on constrained surface coated with Poly(Bisphenol A carbonate) of molecular weight 45,000 $g/mol$.  (a) 2D image of pore patterns (scale: 0-250 $\mu m$) (b) zoomed view (c) 3D image of the zoomed portion (scale: x = $0-550$ $\mu m$; y = $0-550$ $\mu m$; z = $0-20$ $\mu m$).}
\label{pc45const}
\end{figure}

\begin{figure}[!h]
\centerline{\includegraphics[width=5.5in]{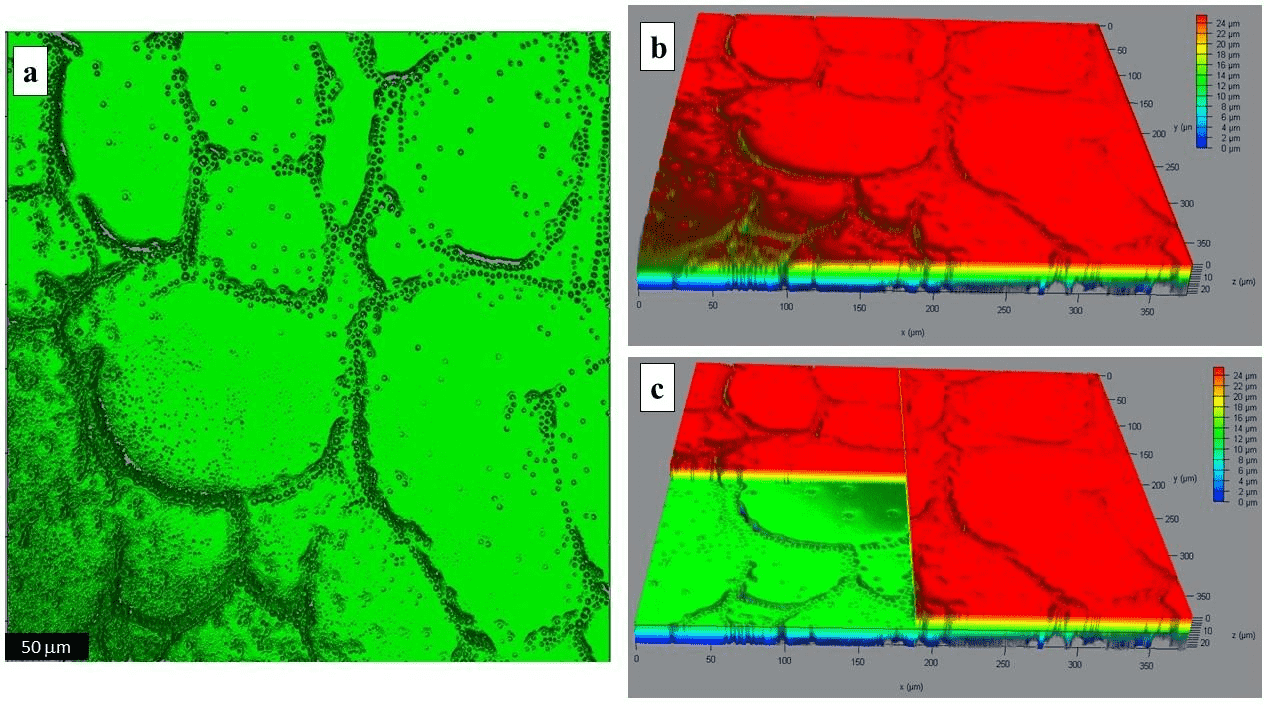}}
\caption{CLSM image of pattern formation on constrained surface coated with Poly(Bisphenol A carbonate) of molecular weight 28,200 $g/mol$.  (a) 2D image of pore patterns (scale: $50 \mu m$) (b) 3D CLSM image (scale: x = $0-400$ $\mu m$; y = $0-400$ $\mu m$; z = $0-24$ $\mu m$) (c) 3D sliced image (scale: x = $0-400$ $\mu m$; y = $0-400$ $\mu m$; z = $0-24$ $\mu m$).}
\label{pc28const3d}
\end{figure}

The formation of self-assembled droplet patterns are influenced by many factors.  Type of polymer and the molecular weight of polymer are two among them.  To study the influence of former, two different polymers: PDMS and Poly(Bisphenol A carbonate) are used. To evaluate the influence of latter, two different molecular weights of Poly(Bisphenol A carbonate) are used. The confocal images of patterns formed on polycarbonate films over constrained surfaces are shown in Figures ~\ref{pc45const} and ~\ref{pc28const3d}.  Both the images demonstrate very similar morphology to that of PDMS.  The differences in pore sizes and ring size are observed and are given in the Table~\ref{tab:1}. From the Table~\ref{tab:1}, it is observed that the pore size and ring size increases with increase in molecular weight of the polymer.  \\
\begin{table*}
\small
\caption{Average pore parameters for constrained substrates patterned with different polymers (Error:$\pm 0.1 \mu m$)}
\label{tab:1}
\centering
\begin{tabular*}{\textwidth}{@{\extracolsep{\fill}}l|c|c|ccc}
\hline
            & \multicolumn{3}{c}{Pattern formation on constrained surface coated with}   \\  \cline{2-4}  
            
 \multirow{3}{*}{Parameters}           & PDMS       & Poly(Bisphenol A carbonate)  &Poly(Bisphenol A carbonate) \\
						&         & M.Wt: 28,200 $g/mol$           & M.Wt: 45,000 $g/mol$ \\ \hline
Average Size of the pores &            &              &    \\ 
on the ring ($\mu m$)    &  3     &  7   & 13  \\
                         &  &  &  \\
Average Size of the pores  &            &           &   \\
inside the  ring ($\mu m$) &  1 & 1.2  &  2.2 \\
  &  &  &  \\
Average Ring size ($\mu m$) & 82         & 152          &300   \\

\hline
\end{tabular*}
\end{table*}

\newpage{}
\subsection{Pore patterns on various polymer thickness}
\begin{figure}[h]
\centerline{\includegraphics[width=6in]{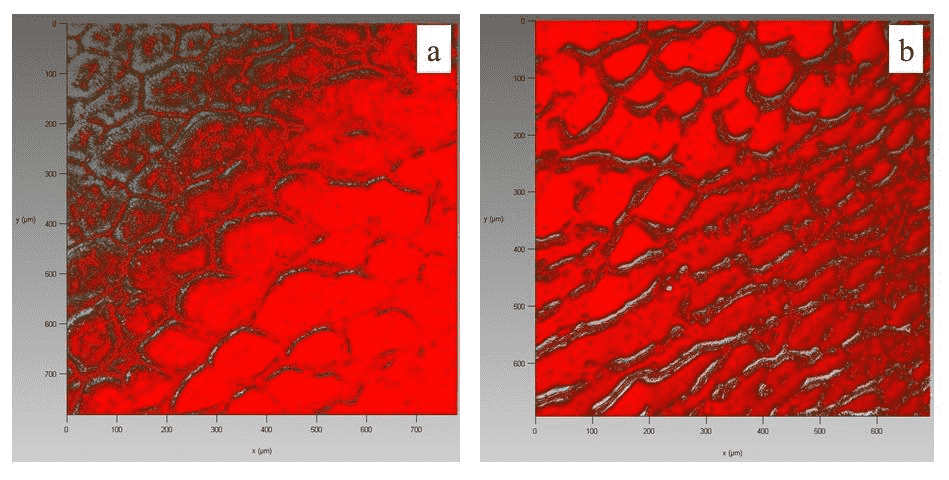}}
\caption{CLSM image of pattern formation on constrained surface coated with PDMS of thickness (a) 17 $\mu m$ (scale x = 0-800 $\mu m$; y = 0-800 $\mu m$) (b) 50 $\mu m$ (scale x = 0-700 $\mu m$; y = 0-700 $\mu m$). }
\label{pdmsthickness}
\end{figure}

To further verify the influence of surface roughness, the droplet patterns were formed over various thickness of polymer film.  It was observed from the patterns formed that the constraints on the underlying surface play a significant role in the formation of the patterns consisting of small scale closely arranged pores within a large scale distorted hexagonal network.  It can be inferred that the the large scale distorted hexagonal network patterns were formed on the constrained surface irrespective of the thickness of the polymer layer up to $50$ $\mu m$. The CLSM images are shown in Figure~\ref{pdmsthickness}. At the thickness of $17$ $\mu m$ the rings are closed, whereas at the thickness of $50$ $\mu m$ the rings are observed to be elongated.   
\subsection{Patterns on smooth surface}
\begin{figure}[h]
\centerline{\includegraphics[width=6in]{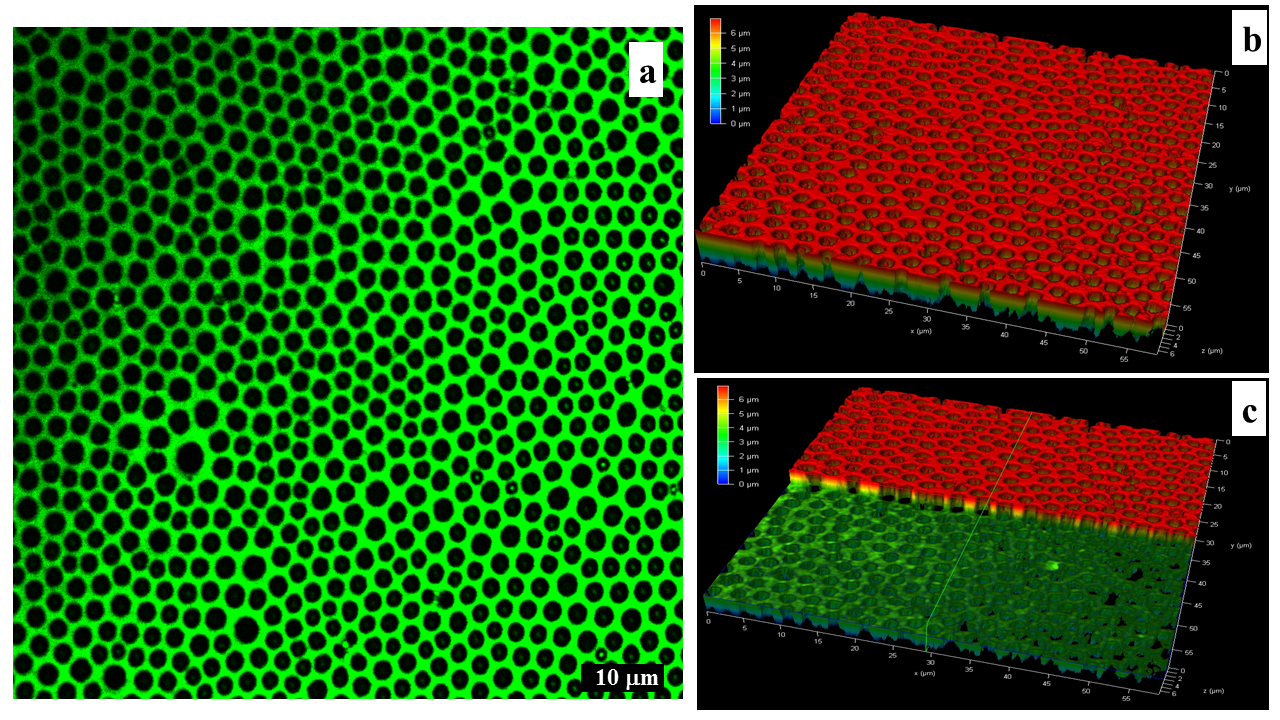}}
\caption{CLSM image of pattern formation on PDMS coated smooth surface  a) 2D CLSM image (scale: $10$ $\mu m$) b) 3D image showing depth (scale: x = $0-60$ $\mu m$; y = $0-60$ $\mu m$; z = $0-6$ $\mu m$) and c) Sliced 3D image.}
\label{pdmssmooth}
\end{figure}

\begin{figure}[h]
\centerline{\includegraphics[width=5in]{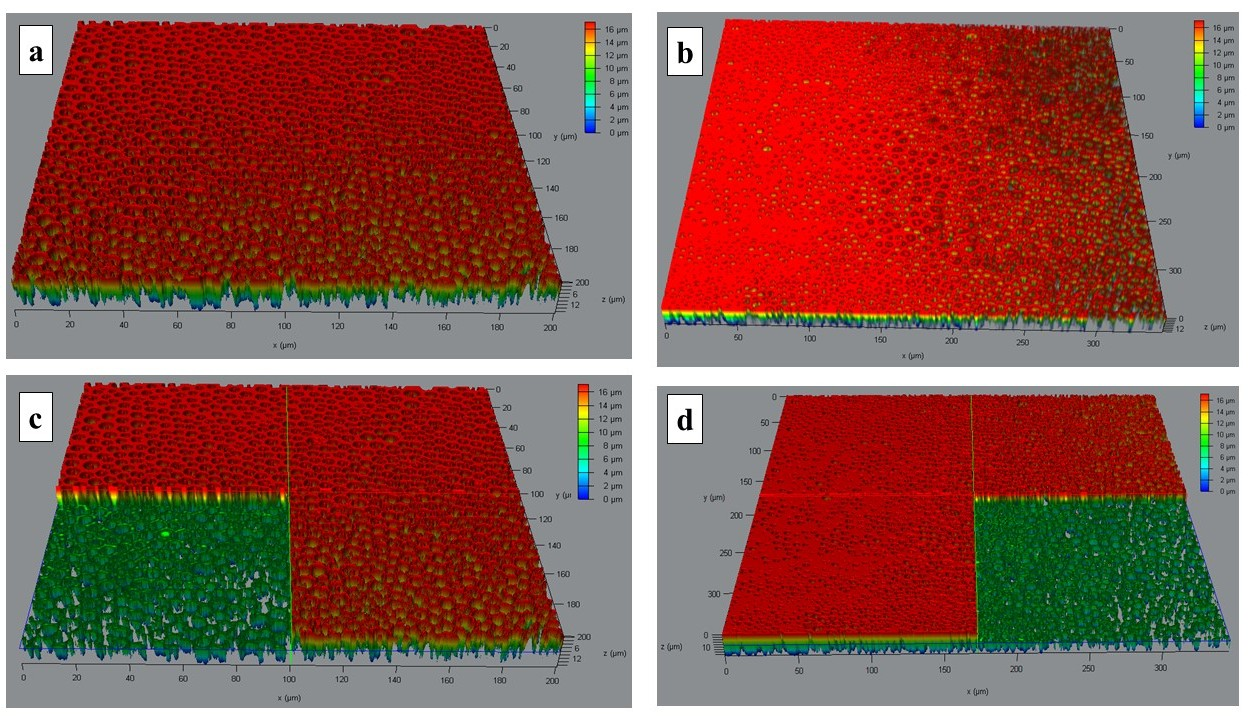}}
\caption{3D image of surfaces patterned with (a)  Poly(Bisphenol A carbonate) of molecular weight 28,200 $g/mol$ (scale: x = $0-200$ $\mu m$; y = $0-200$ $\mu m$; z = $0-16$ $\mu m$) and  (b) Poly(Bisphenol A carbonate) of molecular weight 45,000 $g/mol$ (scale: x = $0-400$ $\mu m$; y = $0-400$ $\mu m$; z = $0-16$ $\mu m$). Sliced view of surfaces patterned with (c) Poly(Bisphenol A carbonate) of molecular weight 28,200 $g/mol$ (scale: x = $0-200$ $\mu m$; y = $0-200$ $\mu m$; z = $0-16$ $\mu m$) and (d) Poly(Bisphenol A carbonate) molecular weight:45,000 $g/mol$ (scale: x = $0-400$ $\mu m$; y = $0-400$ $\mu m$; z = $0-16$ $\mu m$).}  
\label{pcsmoothclsm}
\end{figure}

Comparative studies are made with smooth surfaces and the morphology of the patterned smooth substrate with various polymers are shown in Figure~\ref{pdmssmooth} and ~\ref{pcsmoothclsm}.  Pore patterns on the smooth surface show honey comb pattern with the pore diameter of around $2.5$ $\mu m$. The pores of the honey comb pattern are fairly uniform in size which confirms the unimodal distribution of droplets and also the absence of coalescence of droplets. The honey comb patterns formed on the smooth surface are driven by the conventional surface tension driven Marangoni instability and the presence of lubricating polymer film from instant precipitation of polymer.\\

\subsection{Wetting studies}
\begin{figure}[h]
\centerline{\includegraphics[width=5in]{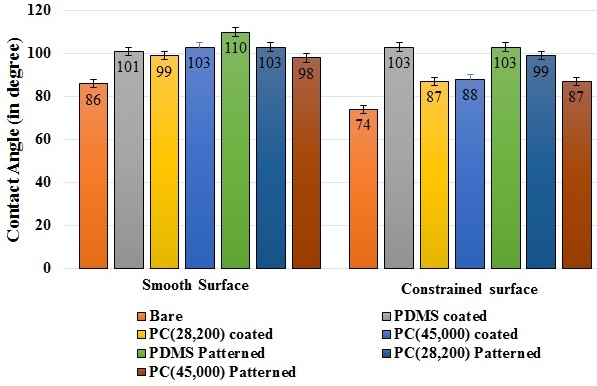}}
\caption{Contact angles of $3$ $\mu l$ of distilled water over the various substrates before and after patterning. In the figure, "bare
smooth/constrained" refers to the polycarbonate substrate without any PDMS/Poly(Bisphenol A Carbonate) coating.}
\label{cacomp}
\end{figure}

The water contact angle of the surfaces before and after patterning are given in Figure~\ref{cacomp}. The bare smooth surface showed the contact angle of $86$$^{o}$, whereas the bare constrained surface showed a decrease in contact angle. This is due to the presence of grooves on the constrained surface which obeys Wenzel's equation ($cos \theta^{\ast} = R \cos \theta$; where $R$ is the roughness factor, $\theta^{\ast}$ is the Wenzel angle and $\theta$ is the Young's angle). According to Wenzel's equation, the increase in hydrophobicity/hydrophilicity increases with increase in surface roughness.  Since the bare polycarbonate smooth surface is hydrophilic, its contact angle further decreased with the presence of grooves (roughness). \\

\begin{figure}[!h]
\centerline{\includegraphics[width=5in]{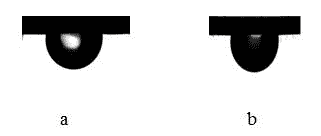}}
\caption{Representative images showing high adhesive force of surface to water: a) Smooth surface patterned with PDMS
 and b) Constrained surface patterned with PDMS with droplet turned upside down.}
\label{wen}
\end{figure}

Increase in contact angle is observed for both smooth and constrained surfaces patterned with PDMS and Poly(Bisphenol A carbonate) of molecular weight 28,200 $g/mol$ . These changes in contact angle are attributed to the change in surface roughness resultant from pore formation and also the inter-facial energy of the surfaces. However, there is no considerable change in contact angle is observed for surfaces patterned with Poly(Bisphenol A carbonate) of molecular weight 45,000 $g/mol$. This may be due to the formation of larger pores on the ring whose diameter is larger than the water droplet used in the contact angle measurements. This require further theoretical investigations.  The droplets over all patterned surfaces was seen to be in a sticky Wenzel state \cite{wenzel}, as seen from Figure~\ref{wen}.  This is confirmed by upturning the substrate with water droplet and observing the variation of the pendant drop. It is found that the droplet sticks to the surface which confirms the high adhesiveness of droplet with the surface. The droplet pinning is observed to be higher for the smooth surface. The images of the droplets sticking to PDMS patterned smooth and constrained surface are shown in Figure~\ref{wen}.
\newpage{}

\newpage{}
\begin{figure}[h]
\centerline{\includegraphics[width=5in]{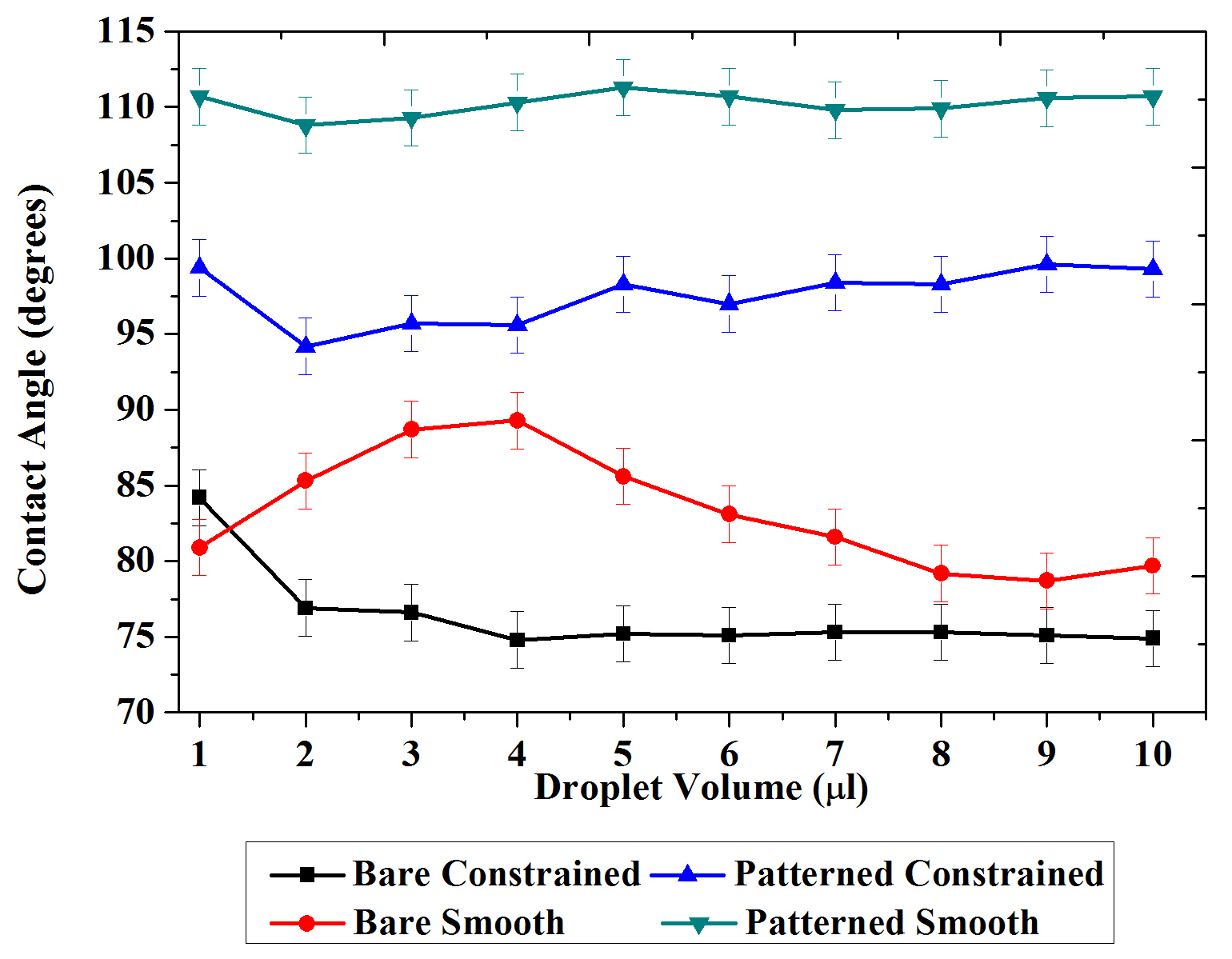}}
\caption{Variation of contact angle with droplet volume for various substrates.}
\label{dropvol}
\end{figure}
\begin{figure}[!h]
\centerline{\includegraphics[width=5in]{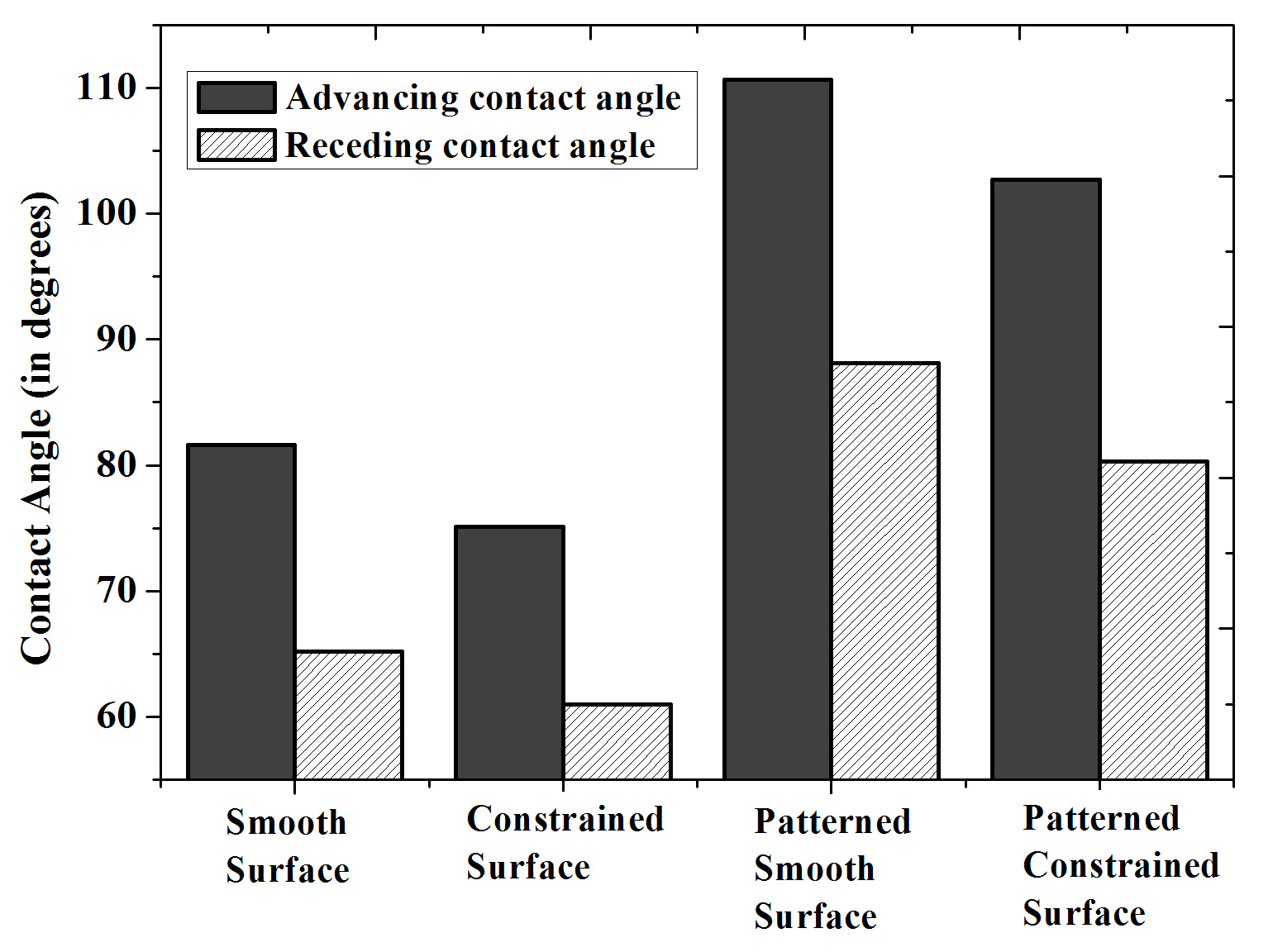}}
\caption{Contact angle hysteresis of various substrates}
\label{cah}
\end{figure}
To further observe whether the surface stays in Wenzel state for increasing droplet volume, the variation of contact angle with the change in droplet radius for all the patterned and unpatterned surfaces are measured and is shown in Figure~\ref{dropvol}. It is observed that there is no considerable variation in contact angle with increasing droplet volume. This confirms that the droplets are sticking to the surface showing Wenzel's state of wetting and is independent of the volume of the liquid droplet. For further confirmation, contact angle hysteresis are performed and all showed a high contact angle hysteresis indicative of Wenzel's state \cite{wenzel}. The experimental results are shown in Figure~\ref{cah}.

\subsection{\textit{Surface Evolver} results}
In the present study, a numerical simulation has been performed to analyze the droplet shape on various substrates namely, smooth and constrained surfaces.  The constrained surfaces are obtained by varying the geometrical parameters of the surface.  After validating the simulation results with the experimentally obtained data, the simulation is extended to various liquids.  \\

\begin{figure}[!h]
\centerline{\includegraphics[width=3in]{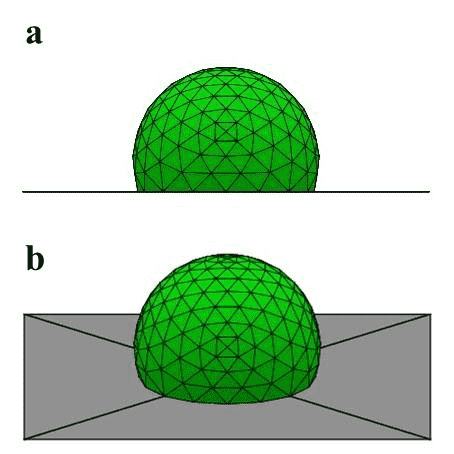}}
\caption{Equilibrium shape of the droplet on smooth surface obtained from \textit{Surface Evolver} a) side view b) upper view}
\label{pdmssmoothse}
\end{figure}
\begin{figure}[h]
\centerline{\includegraphics[width=3in]{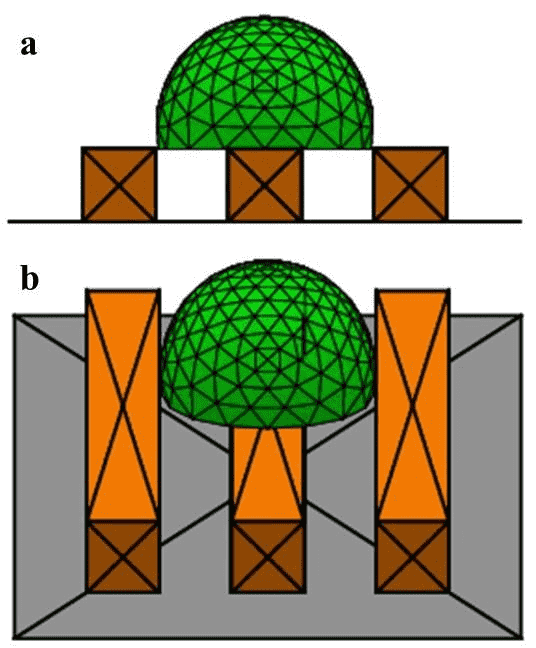}}
\caption{Equilibrium shape of the droplet on constraint surface obtained from \textit{Surface Evolver} a) side view b) upper view}
\label{pdmsconstraintse}
\end{figure}

\begin{figure}[h]
\centerline{\includegraphics[width=4in]{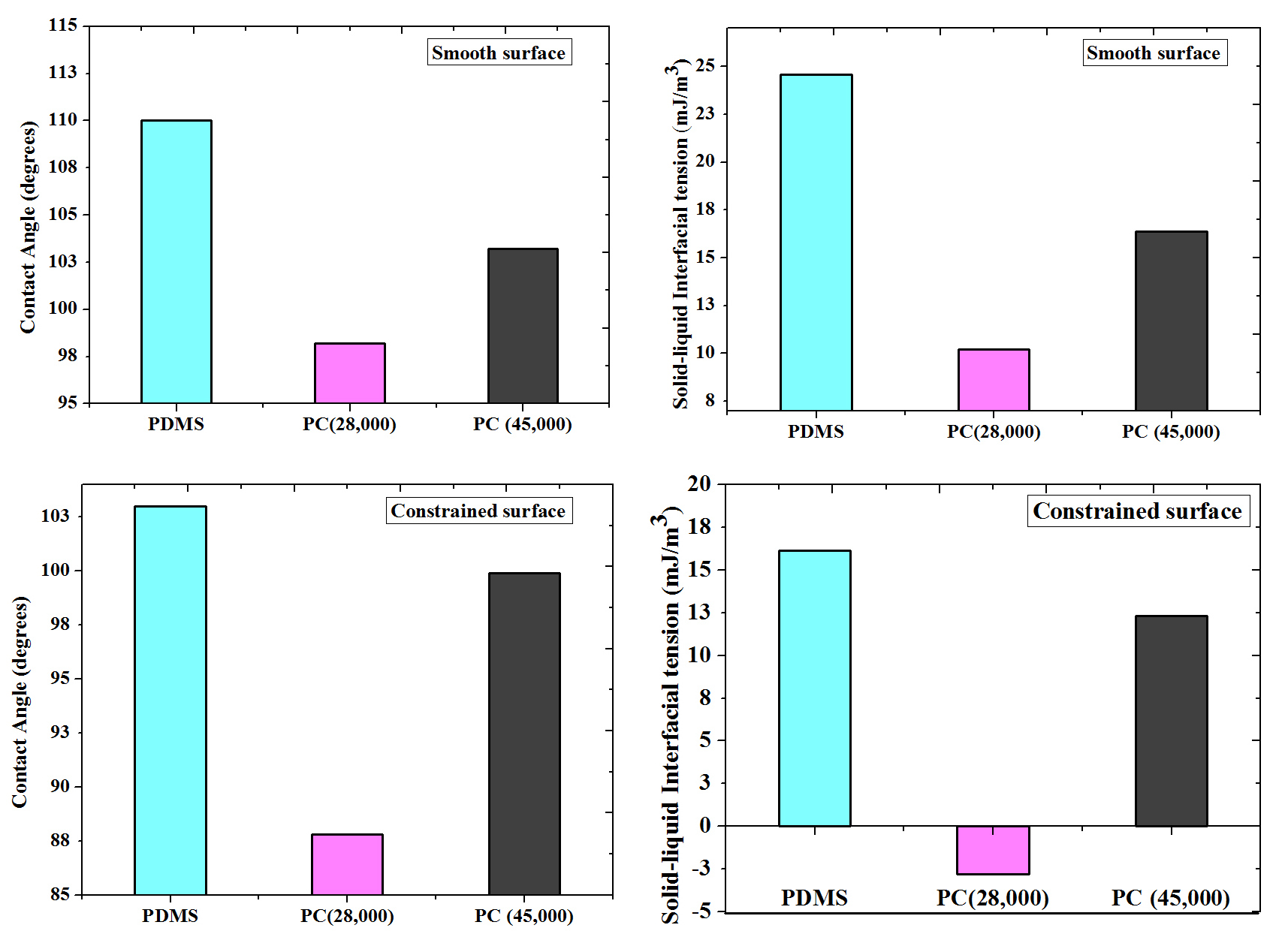}}
\caption{Contact angle of water on surfaces with various polymer solution and solid-liquid inter-facial energy of various polymers with water (Top-left : Contact angle of water on smooth surface; bottom-left: Contact angle of water on constrained surfaces; top-right: solid-liquid inter-facial energy of smooth surface and bottom-right: solid-liquid inter-facial energy of constrained surface) }
\label{anglese}
\end{figure}

For the present work, a polycarbonate sample with three microgrooves of constant groove width and depth are selected as constrained surface for drop shape analysis.  Constant drop volume of $9$ $\mu l$ is maintained through out the study.  The simulated drop shapes of water on PDMS coated smooth and constrained surfaces are shown in the Figure~\ref{pdmssmoothse} and ~\ref{pdmsconstraintse}. The primary objective of performing the \textit{Surface Evolver} simulation is to examine the effect of roughness geometry on the wettability of a surface.  The key parameter in analyzing the wetting behavior of a surface is its apparent contact angle.  The contact angle is found to be influenced by the solid-liquid inter-facial tension. The inter-facial energy of various polymers and the contact angle of water on different surfaces considered in the present study are shown in Figure~\ref{anglese}. \\ 

\begin{figure}[h]
\centerline{\includegraphics[width=4in]{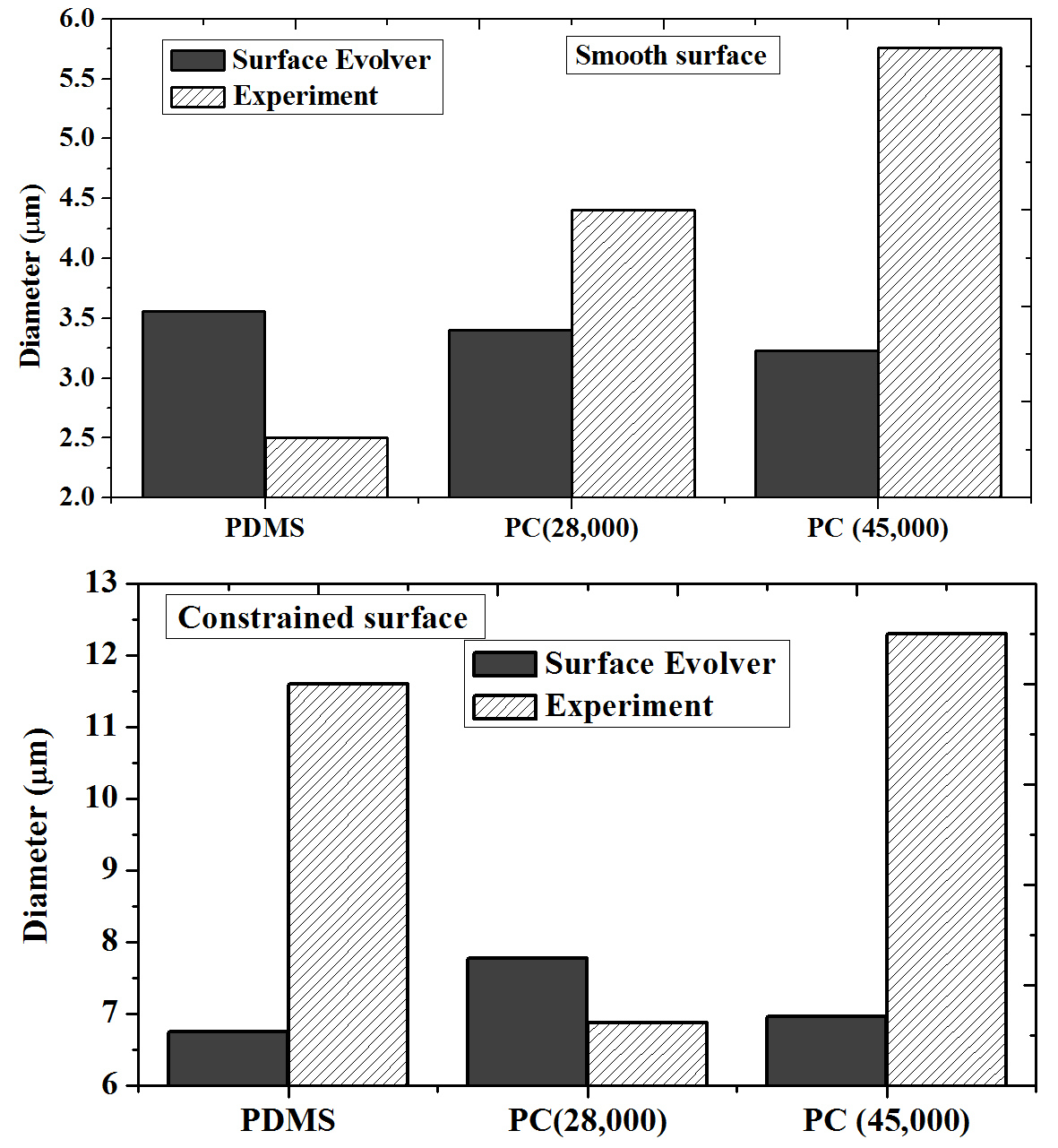}}
\caption{Comparison of droplet diameter between experimental and simulation results. (Top: smooth surface; Bottom: Constrained surface)}
\label{diameterse}
\end{figure}

\begin{figure}[!h]
\centerline{\includegraphics[width=4in]{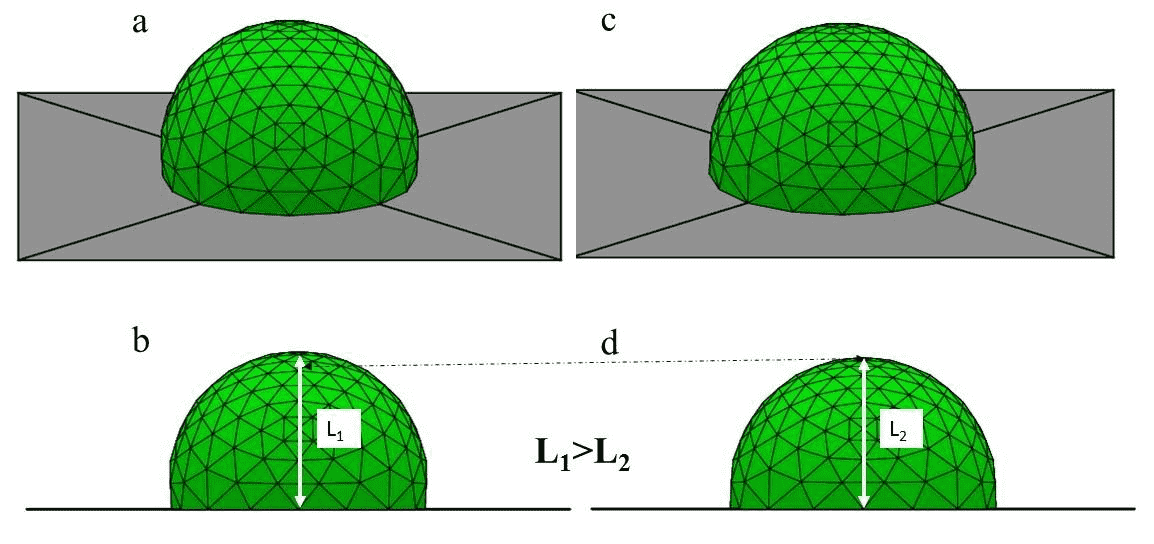}}
\caption{Comparison of equilibrium drop shape of water on smooth surfaces with various polymer solution (a) Poly(Bisphenol A carbonate)-M.wt:45,000 - upper view b) Poly(Bisphenol A carbonate)-M.wt:45,000 $g/mol$ - side view  c) Poly(Bisphenol A carbonate)-M.wt:28,000 $g/mol$ - upper view  and d) Poly(Bisphenol A carbonate)-M.wt:28,000 $g/mol$ - side view.}
\label{pcsmoothse}
\end{figure}

\begin{figure}[!h]
\centerline{\includegraphics[width=4in]{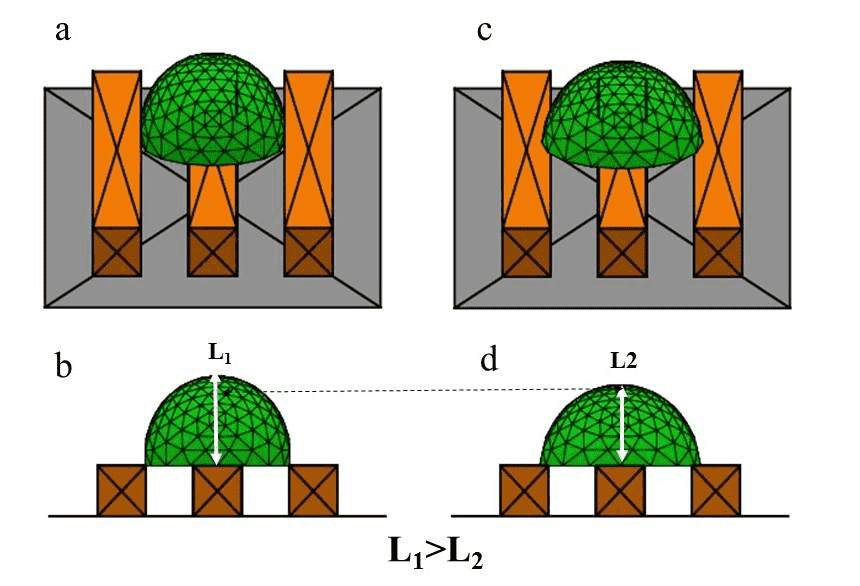}}
\caption{Comparison of equilibrium drop shape of water on constrained surfaces with various polymer solution (a) Poly(Bisphenol A carbonate)-M.wt:45,000 $g/mol$- upper view b) Poly(Bisphenol A carbonate)-M.wt:45,000 $g/mol$ - side view  c) Poly(Bisphenol A carbonate)-M.wt:28,000 $g/mol$ - upper view  and d) Poly(Bisphenol A carbonate)-M.wt:28,000 $g/mol$ - side view.}
\label{pccompconse}
\end{figure}

\begin{figure}[!h]
\centerline{\includegraphics[width=4in]{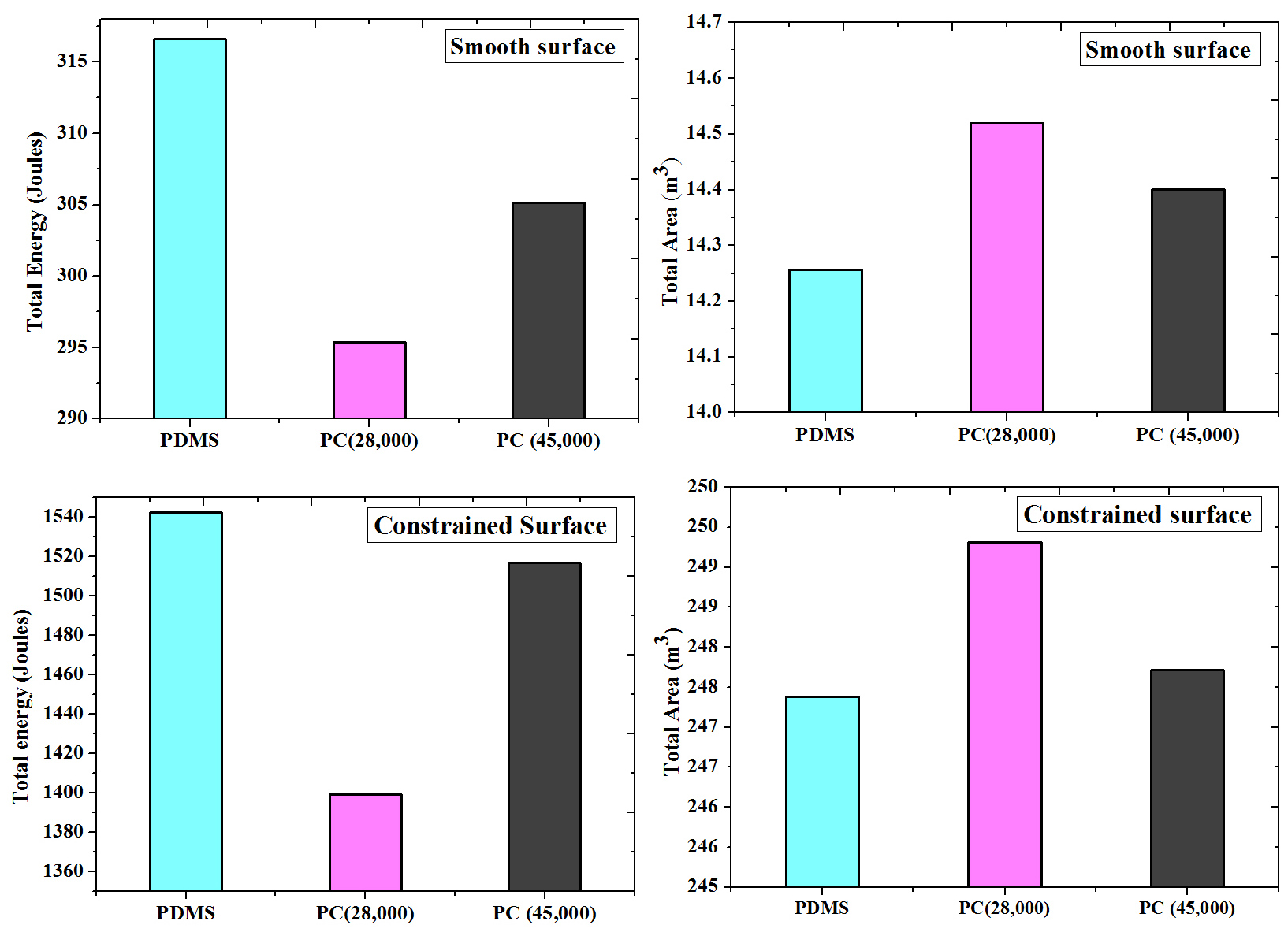}}
\caption{Comparison of total energy and total area of water droplets on various polymers obtained from \textit{Surface Evolver}. (Top-left : total energy of smooth surface; bottom-left: total energy of constrained surface; top-right: total area of smooth surface and bottom-right: total area of constrained surface). }
\label{energyse}
\end{figure}

To investigate the wetting behavior of smooth and constrained surfaces, the numerical results and the experimental findings are compared and the results are shown in Figure~\ref{diameterse}. From the figures it can be clearly seen that the simulations and experimental results show a similar trend.  The deviations at certain points are primarily due to the presence of chloroform in the ambient experimental conditions which is not taken into account in the simulations.   \\
The equilibrium shape of the water droplet on Poly(Bisphenol A carbonate) coated smooth surface is shown in Figure~\ref{pcsmoothse}. It is observed from the figure that the height of the droplet is higher for Poly(Bisphenol A carbonate) of molecular weight 45,000 $g/mol$.  \\ 
Comparison of equilibrium drop shape of water on constrained surfaces coated with various molecular weight Poly(Bisphenol A carbonate) solution are shown in Figure~\ref{pccompconse}. It is observed that the height of the droplet is higher for polymer of higher molecular weight. The total area and energy of the systems (droplet over smooth and constrained substrate) after attaining equilibrium are plotted and are shown in Figure~\ref{energyse}.  From the figures it is clear that the total area and total energy of both the systems (droplet over smooth and constrained surface) shows an inverse relationship. It can also be concluded that for both smooth and constrained surfaces with Poly(Bisphenol A carbonate), the energy required to reach equilibrium increases with increase in chain length and the equilibrium area decreases for higher chain length. \\ 
	\newpage{}

\section{Conclusion}
The influence of underlying constraints on the self-assembly of liquid droplets was studied by forming self-assembled droplet patterns over a micro-grooved constrained substrate.  Large scale pattern formation consisting of small scale closely arranged droplets inside the large scale distorted ring of droplets was observed on constrained substrates whereas smooth substrates showed the ideal honey comb patterns.  Comparative studies of pore patterns formed on smooth and constraint substrates and the simulated energy minimized shape of the
droplets on smooth and constrained substrates were also discussed.  Patterning was repeated with different polymers the results showed a considerable variation in pore sizes whereas pore patterns are unaffected by variation in polymer. This clearly showed that the underlying constraints have significant role in the formation of large scale pattern formation consisting of small scale closely arranged droplets inside the large scale distorted ring of droplets. Besides all these, the detailed understanding of the influence of grooves and the formation of large scale pattern formation consisting of small scale closely arranged droplets inside the large scale distorted ring of droplets  remains a challenging task \cite{fedorets}.

\newpage{}

\end{document}